\documentstyle[epsf]{elsart}
\input{prepictex}
\input{pictex}
\input{postpictex}
\newdimen\picraise
\newcommand\picbox[1]
{
  \setbox0=\hbox{\input{#1}}
  \picraise=-0.5\ht0
  \advance\picraise by 0.5\dp0
  \advance\picraise by 3pt      
  \hbox{\raise\picraise \box0}
}

\newcommand{\bqn}{\begin{eqnarray}}
\newcommand{\eqn}{\end{eqnarray}}
\newcommand{\nn}{\nonumber\\}

\begin{document}
\begin{frontmatter}
\title{Bound states of massless fermions as a source for new physics}
\normalsize
Opening lecture given at the 33rd course of the International School
of Subnuclear Physics, Erice, Sicily (1995)
\author[itkp]{V. N. Gribov\thanksref{gri}}

\address[itkp]{Landau Institute for Theoretical Physics, ul. Kosygina
                       Moscow, Russia\\
                         and\\
KFKI Research Institute for Particle and Nuclear Physics, H-1525
Budapest 114, P.O.B. 49, Hungary}
\thanks[gri]{E-mail: gribov@itkp.uni-bonn.de}

\begin{abstract}
The contribution of interactions at short and large distances to
particle masses is discussed in the framework of the standard model.
\end{abstract}

\end{frontmatter}

\section{Introduction} \label{I}

       In this opening lecture I would like to discuss the standard
model from a point of view different from the usual discussion of this
theory.

       What is the Weinberg-Salam-Glashow standard model ? We have
three interactions: $SU(3)$, $SU(2)$ and $U(1)$ and three generations
of quarks and leptons. All particles in this theory are supposed to be
intrinsically massless. In order to connect this theory with the real
world, the so-called Higgs sector is introduced, which is responsible
for fermionic masses and the masses of $W^+$, $W^-$ and $Z^0$.

       While the first part of the theory feels beautiful, the second
part of it looks ugly to almost everybody. What is the reason for
that ? The first part - massless gauge bosons and massless fermions -
has several remarkable properties. At the first sight it contains no
dimensional parameters, only dimensionless couplings $g_1$, $g_2$,
$g_3$. But, as it was discovered in 1957 by Landau, Abrikosov and
Khalatnikov \cite{landau}, the Abelian coupling is increasing with momentum,
while the non-Abelian couplings $g_2$, $g_3$, as found by Gross,
Politzer and Wilczek \cite{gross}, are decreasing with it. Experimentally we
know, that \(\alpha_3=\frac{g_3^2}{4\pi}\) is of the order of unity in
the momentum region $\lambda_3\sim1$ GeV and
$\alpha_1=\frac{g_1^2}{4\pi}$ is of the order of unity at the scale
$\lambda_L=10^{38}$ GeV; $g_2$ is small everywhere. As a result, instead
of dimensionless couplings we have here two scales, short and large
distance ones different by $10^{38}$ orders of magnitude. Between these two
scales all interactions are weak.

Introducing the Higgs sector, we are introducing a new scale
due to the running of the Higgs self-coupling $\lambda$ - a scale,
where this $\lambda$ is of the order of unity. If the Higgs boson was
light enough, this new scale would be relatively large. However, no
light Higgs boson was found so far and in this sense the theory of the
trivial Higgs sector is close to being in trouble.

       There were many proposals - supersymmetry, grand unification,
supergravity etc. - which corresponded to the introduction of new
interactions between the two scales I mentioned. Some of them may be
correct - we don't know. But to my knowledge nobody was able yet to
construct a theory which is asymptotically completely free in the
sense that it does not contain strong interactions at short distances
like $U(1)$ interactions.

       The aim of my lecture is to raise the question, whether it is
really necessary to introduce new interactions between those two
scales in order to explain the nature, and to present some observations
which indicate that this might not be necessary.

       We believe, that strong $SU(3)$ interactions are able to
prevent quarks from being observed as real particles and to organize
bound states which we call hadrons. Why then the strong $U(1)$
interaction at the scale $q^2\sim\lambda_L^2$ is unable to give
masses to quarks and leptons and their bound states which
would be the Higgs boson and the longitudinal components of $W^+$,
$W^-$ and $Z^0$ ? It is important to understand that the two phenomena
- the existence of bound states and the generation of masses - are
very strongly connected. If the interaction is strong enough to create
massless bound states, these bound states will condensate and the
scattering of massless fermions on this condensate will result in the
propagation of fermions as massive objects. In this sense the source
for the mass generation is the existence of massless bound states of
massless fermions. What kind of bound states can we expect ? If we have
a doublet of left-handed quarks interacting with right-handed
antiquarks through the $U(1)$ exchange, they could create a massless
spinless doublet. Together with an anti-doublet it will give us four
states which, eventually, become massive bosons - the Higgs boson and
three longitudinal components of $W^+$, $W^-$ and $Z^0$. The last step
occurs at large distances: between the Compton wave length of the
quark and $\lambda_L^{-1}$. In order to understand how this can
happen, let us discuss the space-time structure of these
``intrinsically'' massless bosons. Due to the fact, that these bound
states are created by the strong interactions on the distances $10^{-52}$
cm, one might expect that they will be point-like. However, although
there is very little knowledge about the structure of relativistic
bound states, these expectations do not seem to be correct. Even in
the non-relativistic quantum mechanics this expectation is not true
in general, since the size of a bound state is determined by the
binding energy and not by the radius of the forces. A famous example
for this is the deuteron in the
limit when the radius of nuclear forces is going to zero. In this
limit the proton and the neutron inside the deuteron are always
outside the region where the interaction is different from zero. In
reality, in addition to nuclear forces there exist always long-range
electromagnetic interactions between the proton and the neutron. The
binding energy depends on both interactions and would not be zero even
if the binding energy due to solely nuclear interactions were exactly
zero. This is just an analogue of the mechanism by which a
massless boson acquires a mass. In the latter case the role of
e.m. interactions between the proton and the neutron is played by the
interaction with the condensate and with the colour field. The size of
the new massive state would be determined by the masses of the
constituent fermions.

       These considerations suggest, that the boson masses could be
calculated as functions of fermionic masses, with little
sensitivity to the lack of knowledge about the interaction structure
at the distances of the order of $1/\lambda_L$. In the next section I will
present these calculations. They contain a factor
$\ln \frac{\lambda_L^2}{\lambda_3^2}$ and give reasonable
predictions for masses of the top quark and the Higgs meson. In the
third section of this lecture we shall discuss another topics - the
$\pi$ meson mass. The $\pi$ meson is a bound state of a light $q$ and
a light $\overline{q}$ interacting via the exchange of coloured
gluons, which becomes a massless Goldstone in the limit when the
masses of light quarks tend to zero. But the light quark
participates in both $U(1)$  and $SU(3)$ interactions and thus plays a
role in the creation not only of pions but also of massless Goldstones
which are the source of the longitudinal components of $W$ and $Z$.
This means,
that these two interactions, the long range $SU(3)$ and the short
range $U(1)$ produce two types of zero mass bound states. It
is natural to expect, that these states will strongly influence each
other. It is possible to calculate the interaction between these two
states. As we will show, this gives an expression for the $\pi$ meson
mass as an integral over light quark masses and this integral contains
the same $\ln \frac{\lambda_L^2}{\lambda_3^2}$ as the Higgs
mass. Knowing the $\pi$ meson mass and using the same value for the
Landau scale, we find a reasonable value for the constituent
light quark mass.

       The conclusion of sections 2 and 3 is, that measuring
masses of some bound states of quarks and leptons we obtain
information about the scale where the fermionic masses are created,
and this information is roughly in agreement with the
idea that masses are generated at the Landau scale. If proved to be
true, this would suggest, that the other interactions, even if they
exist between the two scal

es, are not important in the process of
creation of fermion masses.

       Before going to these rather simple calculations, let us discuss,
what kind of problems will arise if one assumes that $U(1)$
interactions are responsible for the fermionic masses. The source of
the problem is obvious: there are many quarks and leptons, so why do we
expect to have only four intrinsically massless bosons ? Indeed, these
bosons are proved to be complicated superpositions of different quarks
and leptons. Where are the other massless states which we have to
expect on the basis of the $U(1)$ symmetry ? A possible answer is the
following. The interaction I call $U(1)$ is well defined only in the
region of momenta much smaller than $\lambda_L$; in the region
$q^2\geq\lambda_L^2$ where the coupling is large we are not able to
write any Lagrangians, since in this region the $U(1)$ interaction
could be considered as a very complicated fermionic interaction or even
an interaction which contains an additional horizontal gauge field
connecting different generations. The only condition imposed on this
interaction is, that it has to have no logarithmic tails into the
$q^2\ll\lambda_L^2$ region.
It is not clear at all, whether the interaction is uniquely
defined by this condition. The strongest objection to a theory of this
type is usually connected with the existence of a large number of
non-anomalous currents which are conserved for massless fermions; for any
of these currents there have to be corresponding Goldstones. This
argument is, however, not so serious as it looks. The contribution of the
Goldstones to the currents is proportional to the product $fg=m$ where
$f$ is the amplitude for the Goldstone current transition, $g$ is
the analogue of the Yukawa coupling and $m$ is the fermion mass. If
$f$ is of the order of
$\lambda_L$, then $g \sim \frac{m}{\lambda_L}$ and the Goldstone will
not interact with fermions.

\section {Masses of vector bosons and Higgs bosons}

       Let us consider the polarization operator for $SU(2)$ vector
bosons. It can be written as
\begin{equation}
  \label{e1}
  \Pi_{\mu \nu}(k) = \,\, \input{gribov1.pictex}
\end{equation}
with vertices in $\Pi_{\mu\nu}$ of the form
$\frac{1}{2}\tau_\alpha\frac{1}{2}(1-\gamma_5)\gamma_\mu$. For
massless fermions $\Pi_{\mu\nu}$ has the form
$\Pi_{\mu \nu} = \Pi_{\mu\nu}^\perp = (\delta_{\mu\nu}k^2-k_\mu
  k_\nu) \, \Pi (k^2)$, and it corresponds to the massless $W$. Fermionic
masses lead to an additional term $\Pi_{\mu \nu}'$ in
$\Pi_{\mu \nu}$:
\bqn \label{e2}
\Pi_{\mu \nu}' & = & \,\, \input{gribov2.pictex} \nn
& = & \delta_{\mu\nu} \frac{3}{32\pi^2}
\sum_{i} \int_{q^2 > m^2}
\frac{dq^2}{q^2} \, [m_{i \uparrow}^2(q^2)+m_{i \downarrow}^2(q^2)+
\frac{1}{3}m_{i l}^2(q^2) ] \; ,
\eqn
where $\sum_i$ in (\ref{e2}) stands for the summation over all
generations of up ($\uparrow$) and down ($\downarrow$) quarks and
leptons ($l$). Correspondingly, the square of the $W$ mass is equal
\bqn \label{e3}
m_W^2 & = & g_2^2 \, \frac{3}{32 \pi^2} \, \int_{m_t^2}^\infty
\, \frac{dq^2}{q^2}
\, m_t^2(q^2) \; .
\eqn
Here $m_t$ is the top-quark mass, if we keep only the contribution
of the heaviest quarks. The behaviour of $m_t^2(q^2)$ is well-known in
the region where all couplings are small:
\bqn \label{e4}
m_t^2(q^2) = m_t^2 \left[ \frac{g_3(q^2)}{g_3(m_t^2)} \right]^{8/7}
\left[ \frac{g_1(m_t^2)}{g_1(q^2)} \right]^{1/10} \; .
\eqn
According to (\ref{e3}), $m_t^2(q^2)$ changes with $q^2$ very slowly,
and therefore, roughly,
\bqn \label{e5}
m_W^2 \approx
g_2^2 \, \frac{3}{32\pi^2} \, m_t^2 \,
\ln\frac{\lambda_L^2}{m_t^2} \; .
\eqn
In spite of the fantastically large value of the Landau scale, we have
\bqn \label{e6}
\frac{1}{16\pi^2} \, \ln\frac{\lambda_L^2}{m_t^2} \sim 1
\eqn
and thus $m_W \sim m_t$. Taking into account the explicit
dependence of $m_t^2(q^2)$ on $q^2$ according to (\ref{e4}), we find
\cite{gribov}
\bqn \label{e7}
m_W^2 & = & \frac{3}{2} \, \frac{g_2^2(m_t^2)}{g_3^2(m_t^2)}
\, \left\{ 1 - \left[ \frac{g_3^2(\lambda_L^2)}{g_3^2(m_t^2)}
   \right]^{1/7}
   \right\} \; .
\eqn
The accuracy of this result depends on the unknown contribution of the
region $q^2 \geq \lambda_L^2$ and on the strong interaction in the
region $q^2 \leq m_t^2$ as well as on the effective Yukawa couplings
which appear in the theory due to non-zero quark masses.

Because of the smallness of $\frac{1}{16 \, \pi^2}$, which was compensated
by the large value of $\ln \frac{\lambda_L^2}{m_t^2}$, the corrections
coming from the region $q_t^2 \leq m_t^2$ are small. Corrections from
the region $q^2 \geq \lambda_L^2$ will also be small provided $m_t^2(q^2)$
starts to decrease faster (not logarithmically) at $q^2 > \lambda_L^2$.
In the work which was done together with Yu. Dokshitzer
\cite{dokshitzer} we have
estimated the possible value of $m_t$ from the known value of $m_W$
by taking into account contributions of the Yukawa couplings,
reasonable cut-off values in the region $q^2 > \lambda^2$ and
different values for $g_3(m_t)$. We obtained the interval
\bqn \label{e8}
180 \; \mbox{GeV} < m_t < 200 \; \mbox{GeV}
\eqn
for $m_t$. If fermionic masses were created at a scale smaller than
the Landau scale, the resulting top quark mass would, respectively, be
larger.

Written in the form (\ref{e2}), the polarization operator
$\Pi_{\mu\nu}$ for $W$ is not transverse. In order to make it
transverse, we have to add to
the expression (\ref{e1}) the Goldstone contribution:
\begin{equation}
  \label{e9}
  \Pi_{\mu\nu} = \,\, \input{gribov9a.pictex} - \,\,\,
                      \input{gribov9b.pictex}
\end{equation}
Strictly speaking, it is necessary to have three Goldstone states in
(\ref{e9}), and in this sense these three massless states are responsible for
the masses of $W$ and $Z$. At short distances, where the masses of
fermions are small, the existence of three massless bosons implies, that
there has to be a fourth massless state corresponding to a scalar
boson, which, however, could become massive due to long distance interactions;
this state is called the Higgs boson. In terms of the mass of the
heaviest quark (the top quark) the mass squared of the Higgs boson is
equal \cite{gribov}
\bqn \label{e10}
m_H^2 = 4 \; \frac{\int m_t^4 \, \frac{dq^2}{q^2}}
                  {\int m_t^2 \, \frac{dq^2}{q^2}} \; .
\eqn
In the approximation which leads to the restriction (\ref{e8}),
expression (\ref{e10}) gives the Higgs mass in the same range
\bqn \label{e11}
180 \; \mbox{GeV} < m_H < 200 \; \mbox{GeV}  \; .
\eqn

\section{The $\pi$ meson mass}

In order to see how the mass of the $\pi$ meson could be influenced by
the interaction which is responsible for the generation of fermionic masses, we
can start with an $SU(2)$ current. Due to the Goldstone
contribution
\begin{equation}
  \label{e12}
\Gamma_\mu^\alpha = \,\, \input{gribov12a.pictex}
                  - \,\,\, \input{gribov12b.pictex}
\end{equation}
to this current (where $m$ is the quark mass), it has to be conserved for
any fermionic states, and for the light quark states in
particular. For simplicity, we will neglect the mass difference of the
$u$ and $d$ quarks. If we include the $SU(3)$ colour interaction and
suppose, that this interaction produces a bound state called $\pi$,
equation (\ref{e12}) will take the form
\begin{equation}
  \label{e13}
  \input{gribov13a.pictex} = \input{gribov13b.pictex} -
  \input{gribov13c.pictex}
\end{equation}
The current conservation $k_\mu \Gamma_\mu = 0$ implies that on the
pion mass shell we have
\begin{equation}
  \label{e14}
  \input{gribov14a.pictex} \,\,\, = \,\,\, \input{gribov14b.pictex}
\end{equation}
In the first order in light quark masses  $\varphi_\pi$ can be
calculated in the limit of zero quark masses; in this case
 $\varphi_\pi$ is connected with the quark Green's function in a
simple way. It is easy to show, that, if the latter is written in the
form
\bqn \label{e15}
G^{-1}(q) = Z \, (\hat m - \hat q) \; ,
\eqn
then at large $q^2$ we have, in general:
\bqn \label{e16}
\hat m (q) = m(q) + \frac{\nu^3(q)}{q^2} \; ,
\eqn
where $m$ and $\nu^3$ are slowly changing functions of $q^2$; the
zero quark mass limit means $m = 0$. In this case $\pi$ is the
Goldstone and its coupling to fermions (i.e. its wave function) is
\bqn \label{e17}
\varphi_\pi = \frac{Z}{f} \, \frac{\nu^3}{q^2} \, \gamma_5 \; .
\eqn
Combined with the expressions (\ref{e15})-(\ref{e17}) the equation (\ref{e14})
leads to
\bqn \label{e18}
k^2 f_\pi^2 = \frac{12}{16 \pi^2} \int \frac{dq^2}{q^2} m \nu^3 \; ,
\eqn
where the integration goes over the region $q^2 > \nu^2$ or $q^2 >
\lambda_{QCD}^2$ for which (\ref{e16}) is correct; $k^2$ is the square
of the pion mass and $f_\pi \approx 93 \; \mbox{MeV}$. The
right-hand-side of (\ref{e18}) has the same structure as the
expression (\ref{e3}) for the W mass: for slowly varying $m$ and $\nu$
values it is proportional to $\ln \frac{\lambda_L^2}{\lambda_{QCD}^2}$
and this large ln factor is compensated by $16 \pi^2$. The form (\ref{e18})
shows explicitly, how the pion mass feels the scale where the quark
masses are created. One can show, that as a function of the strong
coupling $g_3$ the product $m \nu^3$ behaves in the following way:
\bqn \label{e19}
m(q) \nu^3(q) = m_0 \nu_0^3 \left( \frac{g_3(q)}{g_{3 0}}
\right)^{8/21} \; ;
\eqn
here $m_0$, $\nu_0$ and $g_{3 0}$ are the values of the quantities $m$,
$\nu$, $g_3$ at the minimal $q^2$ where (\ref{e16}) is correct.
($q^2 \approx \nu_0^2$). Using (\ref{e18}) and (\ref{e19}), we will
find  $m_\pi^2$ in the form
\bqn \label{e20}
m_\pi^2 = \frac{12 m_0 \nu_0^3}{16 \pi^2 f_\pi^2}
\ln \frac{\lambda_L^2}{\lambda_0^2} \;
\left(\frac{g_3(\lambda_L^2)}{g_{3 0}} \right)^{8/21} \; .
\eqn
Since we do not know $\nu$, the value of $m_\pi$ cannot be checked
yet. However, being the part of the quark Green's function, it has to
enter another physical quantity, like the $\rho$ meson mass and
therefore it can be measured. Calculating $\nu_0$ by making use of (\ref{e20})
and a reasonable value for $g_{3 0}^2$, we find
\bqn \label{e20a}
230 \; \mbox{MeV} < \nu < 260 \; \mbox{MeV}
\eqn
and $1/2 < \alpha_3(\nu) < 1$. This value of $\nu$ is in excellent
qualitative agreement with experiment, because, according to
(\ref{e15}), (\ref{e16}), $\nu$ is the mass of the constituent quark
($G^{-1}(\hat q = \nu) = 0$ if $m_0 \ll \nu$). The formula (\ref{e18})
for $m_\pi^2$ is very close to the widely used expression for the
$\pi$ meson mass
\bqn \label{e21}
m_\pi^2 = \frac{2 \, m_q}{f_\pi^2} <\overline \Psi \Psi> \; .
\eqn
Indeed, (\ref{e21}) can be obtained from (\ref{e18}), if we forget
about $Z$ in (\ref{e15}) and about the running of $m$.

\end{document}